\documentclass{article}

\usepackage{arxiv}

\usepackage[utf8]{inputenc} 
\usepackage[T1]{fontenc}    
\usepackage{hyperref}       
\usepackage{url}            
\usepackage{booktabs}       
\usepackage{amsfonts}       
\usepackage{nicefrac}       
\usepackage{microtype}      
\usepackage{lipsum}		
\usepackage{graphicx}
\usepackage{natbib}
\usepackage{doi}

\usepackage{textcomp}
\usepackage{mathptmx}
\usepackage{float}
\usepackage{graphicx}
\usepackage{soul}\setuldepth{article}
\usepackage{caption}
\usepackage{subcaption}
\usepackage[T1]{fontenc}
\usepackage[scaled=0.85]{beramono}
\usepackage{listings}
\def\niso{\textsc{NIS Ontology}}
\def\hb{\hbox to 11.5 cm{}}

\begin{document}


\title{An Ontological Approach to Compliance Verification of the NIS 2 Directive}

\author{ {\hspace{1mm}Giampaolo Bella}\\
	Department of Mathematics and Computer Science,\\ University of Catania,\\ Viale Andrea Doria 6 - 95125 - Catania, Italy \\
	\texttt{giamp@unict.it} \\
	\And
	{\hspace{1mm}Gianpietro Castiglione} \\
	Department of Mathematics and Computer Science,\\ University of Catania,\\ Viale Andrea Doria 6 - 95125 - Catania, Italy
 \\
 \texttt{gianpietro.castiglione@phd.unict.it} \\
 \And
	{\hspace{1mm}Daniele Francesco Santamaria} \\
	Department of Mathematics and Computer Science,\\ University of Catania,\\ Viale Andrea Doria 6 - 95125 - Catania, Italy\\\texttt{daniele.santamaria@unict.it}}

\date{}

\maketitle

\begin{abstract}
Cybersecurity, which notoriously concerns both human and technological aspects, is becoming more and more regulated by a number of textual documents spanning several pages, such as the European GDPR Regulation and the NIS Directive. 
This paper introduces an approach that leverages techniques of semantic representation and reasoning, hence an ontological approach, towards the compliance check with the security measures that textual documents prescribe.
We choose the ontology instrument to achieve two fundamental objectives: domain modelling and resource interrogation. The formalisation of \textit{entities} and \textit{relations} from the directive, and the consequent improved structuring with respect to sheer prose is dramatically helpful for any organisation through the hard task of compliance verification.
The semantic approach is demonstrated with two articles of the new European NIS 2 directive. 
\end{abstract}


\section{Introduction}
The increasingly rapid growth and complexity of security issues concerns both private and public organisations. 
It could be argued that the broad scope of security measures is demonstrated by recent \textit{security directives}, and a remarkable example of this is European. In 2016, the European Parliament approved the first regulation on security, named the NIS directive, whose recipients are the nations belonging to the European jurisdiction. The directive is a document written without technicalities in natural language containing measures to which the member states must comply. In the last months of 2022, an updated version called NIS 2 totalling 73 pages \cite{NIS} replaced the former in light of the new challenges. In front of this and similar documents of similar size and complexity, this paper faces the overarching problem of how to simplify their reading and make it actionable towards compliance verification.
This is challenging for a variety of reasons due to informal, incomplete and ambiguous prose on one hand, and to linguistic complexity on the other. For example, Alotaibi et al. review such challenges~\cite{inproceedingsPolicies} and OWASP names \textit{``Non-transparent Policies, Terms and Conditions''}~\cite{Owasp} as a privacy risk.
To practically address that problem, this paper sets two research questions:

\begin{quote}
\textit{RQ1: Can we mechanically interpret the relevant entities and relations from security directives given as large documents?}
\end{quote}
Security directives and similar documents are complicated from a linguistic standpoint. If the fundamental actors could be interpreted following a specific pattern, it would be easier to identify their relations and consequently derive an understanding of the general picture as well as of possible pitfalls.

\begin{quote}
\textit{RQ2: Can we mathematically prove that the relations interpreted from security directives hold for specific actors?}
\end{quote}
An example relation is between member states and CSIRT  (Computer Security Incident Response Team) actors; each member state has the specific task to designate its own CSIRT. 
Therefore, answering this research question for a specific member state would mean establishing whether that state designated its own CSIRT.

Formal methods are among the most powerful tools for proving various properties, such as the correctness of security protocols. They take a specific model of the target system and assert, with mathematical support, if the model follows predetermined behaviours. One of the most remarkable formal methods is the inductive one \cite{Bella2000InductiveVO}. We argue that a similar, mathematically rooted approach could strengthen the process of compliance verification with security measures, and follow this argument by advancing a formalisation, which we call \textit{characterisation}, of the NIS 2 directive into the domain of ontologies. 
While the current version of the ontology is a work-in-progress that only covers some excerpts from articles 7 and 10, its benefits in terms of simplified consultation, interpretation and interrogation, thereby providing an answer to \textit{RQ1}.

The ontology, supported by a prototypical implementation for demonstration purposes, allows us to conduct a detailed analysis of the text to identify the semantic patterns of the structures and of the sentences, following the ontology patterns defined by Gangemi et al \cite{inbook}. We give particular attention to the compliance aspect and appeal to logical derivation through inferences and then to queries by description logic, thereby providing an answer to \textit{RQ2}. 

The ontology is released as open-source through a public repository \cite{nisOn}. It must be remarked that its development is currently ongoing towards a more complete coverage of the NIS 2 directive. However, its current version  supports the case that semantic representation provides an effective method for compliance verification.

\section{Paper summary}
The paper is organised in the following way. Section  \ref{sec:related} illustrates the related work on the use of ontologies in the fields of security and its regulations. Section  \ref{sec:overview} introduces the NIS 2 directive. In Section  \ref{sec:model}, we present a high-level representation of the directive, which grounds our approach. In Section  \ref{sec:method}, we specify how  entities and relations are interpreted from the directive, while Section \ref{sec:onto} shows how an  ontology is built upon it, also taking into account the compliance aspect. Section  \ref{Conclusions} summarises our results and outlines the future perspectives.

\section{Related work}\label{sec:related}
The use of ontologies is particularly consolidated outside the security realm, and coherently to our work, also in the legal and compliance context. For example, ontologies were utilised for creating a shared conceptualisation for compliance management \cite{10.1007/978-3-642-31095-9_28}, \cite{4384013}. 
However, as a powerful relational tool, it is increasingly used for security purposes, for example, for modelling smart contracts in blockchains \cite{inbookErc}. 

Ontologies are useful for the classification of generic assets, both hardware and software, of a system or to classify common knowledge, for example, public databases   (CWE - Common Weakness Enumeration, etc.) of vulnerabilities useful for penetration testing \cite{10.1145/1533057.1533084}, \cite{10.1145/2799979.2799995}. 

Another example that deserves a mention is the knowledge base developed by MITRE \cite{mitre} for threat modelling and attack tactics. 

As the reasoning mechanism is peculiar to ontological languages, its usage to track relations and \textit{compliance} is proven. The closest work of structuring security documents, with which we can compare ours, was proposed by Fenz et al. \cite{Fenz2018} with a different target and method. Standards and directives may appear similar in treatment since they both possess a specific structure that defines the order of security measures but with different scopes. Whereas a security standard applies in an industrial context, a security directive relates to political and institutional organisations. This could result in different management of documents and, consequently, in their ontological representation.

Different scope means different recipients which imply different needs. There are two more strong points that allow us to differentiate them, namely, \textit{time} and \textit{possibility}. A security standard might be adopted by an enterprise in an indefinite time at the discretion of the company. Potentially no one could adopt it. Meanwhile, a security directive must be adopted by political institutions at a definite time. Having a tool that speeds up the compliance verification of the directives could be more impactful and in particular if the object has a continental range.

The same cited work superficially treats the method applied for sentence enucleation, focusing more on the implementation part. This paper is focused on that, as we consider the topic fundamental to propose an efficient mechanism for security measures representation.

\section{Description of the NIS 2 directive}\label{sec:overview}
The NIS 2, Directive   (EU) 2022/2555 \cite{NIS}, was adopted on 14 December 2022, and is effective from 16 January 2023, replacing Directive   (EU) 2016/1148. The directive sets rules for security risk management across the nations of the European Union regarding the most important sectors. In some cases, it aligns with related legislation and defines how nations have to cooperate through the establishment of specific groups of relevant stakeholders. The member states will have to adopt the new version of the directive within 21 months. 

The ontological representation of the whole document requires demanding work. However, not all sentences need to be translated into ontological language because they often do not express anything that actors have to fulfil. 
It is possible to extract exactly the parts that are relevant for a complete ontological representation, and we argue that such parts contain chapters from 2 through 7, including Article 6 of Chapter 1. We decided to exclude the early part of Chapter 1 because it only makes general considerations on the applicability of the directive. The same goes for the contents of Chapter 8.

\section{Relational modelling of the NIS 2 directive} \label{sec:model}
Our approach starts with a representation of the directive through a relational model. In fact, one of the main issues when facing security directives is their complexity. For example, it is not obvious that all measures concerning a specific entity are in the same place in the document and, consequently, it is complicated to derive all relations pertaining to the entity. The articles contained in a directive are distributed across the logical sectioning, which is not always obvious.

\begin{figure}[ht]
    \hspace*{1.55cm}
    \includegraphics[scale=0.85]{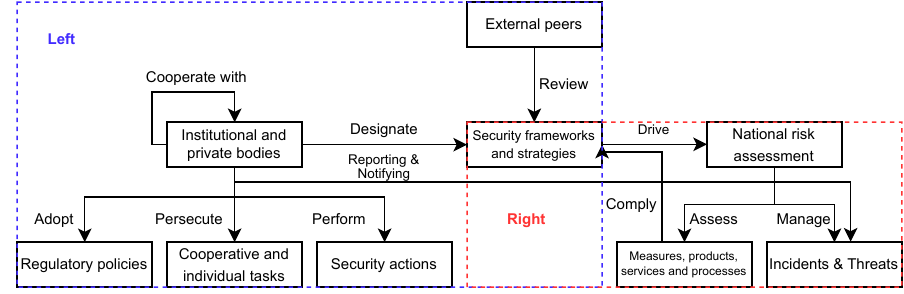}
    \caption{Relational model of entities of the NIS 2 directive}
    \label{fig:kernel}
\end{figure}

We interpreted the relevant entities and relations from the NIS 2 and build a relational model whose nodes contain entities and whose arrows represent relations. 
The model, which is in Figure \ref{fig:kernel}, supports an ontological representation of the directive. For example, it may help the analyst to envisage the role of institutional and private bodies and their interactions, resulting in a valid guide to newbies for the next steps. It allows us to analyse the potential pitfalls and the applicability of each relation. This is useful not only for security analysts but also for the possible future redesigns of the directive.

At this first round, we interpreted entities and relations from the directive by simply reading it, considering the parts of the directive specified in Section \ref{sec:overview}.
The model, from left to right, follows a specific logic, in fact, the specific position of the nodes suggests a first-level sub-modelling. The model we are proposing, in its entirety, could be considered a \textit{Content Ontology Pattern} as defined by Gangemi et al. \cite{inbook}, providing a design solution for the same class of problems.
We appealed to the names of the seven chapters considered, which helped us make a high-level distinction. The two sub-models that can be identified are the left-one and right-one, each of them holding the following properties: 

\subsection{Left sub-model}
We drew the left sub-model referring to articles from 7 to 19, from 26 to 28 and from 31 to 35, each of them defining the role of main bodies and their specific tasks.
On the top, the sub-model has firstly the node `Institutional and private bodies'. In that node, we include all human-composed bodies such as Member States, CSIRT, Eu-Cyclone, which play a primary role in the directive and their eventual composition in groups   (articles 7, 10, 11, 12, 15, 16). The self-referential connection, namely \textit{cooperate with}, makes explicit the concept of cooperation  (articles 13, 14, 17). 
    
The bodies involved in the `Institutional and private bodies' node have different types of tasks: to designate security strategies and to follow assigned and predetermined tasks whose correctness is imposed by specific policies. Each task and policy is defined in the article identifying the related body or in separate articles; this happens, for example, with articles 8 and 9, which refer to a yet-identified body, such as a Member State. Under the definition of tasks, we consider both predetermined and event-driven ones.
The actors, so the various organisations, \textit{persecute} the tasks considered as routine; for example, Article 18 concerns the biennial report of cybersecurity activities. Moreover, we consider the security actions and choose the \textit{perform} relation because this class of tasks has to be performed as a response to events. The same goes for policies: we define as a policy a specific modus operandi to which the bodies must be compliant; therefore, bodies may \textit{adopt} a policy. 

Inside the definition of tasks, we also considered the measures proposed in chapters 7 and 8   (covering articles from 31 to 35) which regard general supervision of entities and jurisdiction aspects.
    
We emphasise the role of `External reviewers' node, which have the task to \textit{review} the security strategies, after their formalisation.
The same idea of disaggregation used previously over tasks is applied here to separate `External peers', whose role is defined in Article 19, from `Institutional and private bodies'. This is sound because `External peers' could be part of external bodies belonging to different organisations.
    
We can recursively repeat the previous considerations regarding tasks, but with a subject that is made of multiple entities. For example, if we  referred to a single CSIRT to perform a task, we can surely agree that the task will also be pursued by the CSIRTs network.
    
It is important to underline that sub-model traversal may not be total. It means that not all entities included in the `institutional and private bodies' node have to fulfil the relational connections. For example, not all of them have to perform security tasks. The main reason for this more-general modelling is just for the sake of simplicity of representation of the relations.

\subsection{Right sub-model}
The right sub-model illustrates the compliance aspects and the role of security strategies, including the activity of information sharing. We took articles from 20 to 25 and from 29 to 30.
The second sub-model is linked to the first sub-model via the \textit{designate} relation. We prefer to exclude the designation from the other tasks because of the node to which it refers. The node representing `Security framework and strategies' expresses all adopted measures, standards, and certifications, which play a central role in a national context, hence it represents the central part of the diagram. The `Security framework and strategies' node encloses citizens, from both private and public points of view, as well as political institutions, thereby acting as a connection point from theoretical, or administrative, activities to practical ones  (article 20). 
The role of this node is to \textit{drive} the activity of risk assessment to \textit{assess} the relevant measures, products, services and processes as well as to \textit{manage} threats and incidents  (articles 21 and 22). 
    
The bodies considered in  `Institutional and private bodies' node must \textit{notify} incidents and threats in time and, possibly, \textit{reporting} voluntarily information related to these incidents and threats   (articles 23, 29, 30).
    
In this sub-model, a node is dedicated to measures, products, services and processes, which together represent the assets meant to be protected by design: they must \textit{comply} with the standards and certification schemes   (articles 24 and 25) expressed by the security strategies promulgated by the aforementioned bodies in the first sub-model. 
The assets that the node represents are grouped together because this is the way they appear in the directive. If we wanted a possible separation we would deduce at the end that the connections with the rest of the model would be the same, for each asset. Therefore, to simplify the model, we can represent them with just one node.

\section{Interpreting the NIS 2 directive}\label{sec:method}

To build an ontological representation of the NIS 2 directive, we first establish how to interpret the constitutional elements of the article and of the measures of the directive. We call this step \textit{semantic interpretation} since it aims to provide meaning to the words of the document towards the ontological representation illustrated in the next Section. 
As previously remarked, the document has some high-level characterising elements: entities and articles. The entities are the minimal elements of the directive as they identify generic assets, both human and technological ones. To establish a criterion to interpret them, we consider the articles as compositions of entities in the form of sentences containing subject, predicate or verb, and object. Those three components indicate who   (the subject) must perform the specific security measure   (the verb/predicate) towards a specific target   (the object). The following sub-sections show how we faced the interpretation of those characterising elements.

\subsection{Interpreting the NIS 2 entities}\label{sec:method:ent}

As a first step, we start with the interpretation of the entities. We can find a list of some of them in Article 6. The entities are not hierarchically organised in such a way as to be useful for ontological representation. For that, we need to establish interpretation criteria according to the peculiarities of the entities. As a first criterion, we can clearly distinguish human-related from technological-related ones.
We also include entities that play a primary role in the directive. For this reason, we establish a  high-level entity named \textit{Actor}. Article 6 does not enumerate all the possible actors and, as a consequence, we need to build a sufficiently general model so as to extend the class hierarchy when required. This step is done while structuring the single relevant article. The entity \textit{Actor} is a glaring example: a CSIRT is not present in the definitions of Article 6 but is clearly an \textit{Actor}. The same applies to the Member State entity and, as we shall see, others. 

Within the \textit{Actor} entity, we can distinguish between human-centred and technological-centred entities, which we respectively identify as \textit{Agent} and \textit{System}. The \textit{Agent} entity includes the possible human-centred entities involved in the security measures as well as those entities that express actions, for example, the \textit{CSIRT}. \textit{Agent} also includes the possible human-centred actors even though they play a passive role in some measures, for example, in the designation of a \textit{CSIRT} by a \textit{Member State}. The reason is that the directive is, by nature, a best practice for someone who must take action to comply. In this phase, we avoid to distinguish an \textit{Agent} further according to its passive and active role.

A different consideration is made for technological-centred entities. The choice of defining role-less entities does not affect the generality of the model. A \textit{System} could be passively used by users and other entities such as a search engine or an online marketplace: they all need to be protected, hence, they are the recipient of a security measure. At the same time, a system could be an active tool. The NIS 2 does not explicitly refer to the tools used in cyber-kill chains during the activities of vulnerability assessment and penetration testing but such activities are very useful to prove the robustness of systems and organisations, the cornerstone of the NIS 2 itself. The same idea is applicable to the entire infrastructure built for security testing. They could be generally identified as \textit{System} and specialised accordingly.

The approach can be straightforwardly applied to other entities. Some of them do not require specific criteria of interpretation: we can identify them as general entities. Other security-related objects can be ousted from the notion of \textit{Actor} and treated accordingly.

\subsection{Interpreting the NIS 2 actors and actions}\label{sec:method:sv}

In this Section, we show how to interpret actors and actions from an article, which is the second characterising element of the directive.

The role of the articles is twofold: on the one hand to define rules that the  subject has to follow and, on the other hand, to further specialise entities. Each article refers at least to a specific agent but complex articles may involve more agents.

For example, we take into account the following excerpt of Article 7:
\textit{``Each Member State shall adopt a national security strategy that provides for the strategic objectives, the resources required to achieve those objectives, and appropriate policy and regulatory measures, with a view to achieving and maintaining a high level of security. The national security strategy shall include... As part of the national security strategy, Member States shall, in particular, adopt policies:...''.}

From the previous excerpt, we deduce that \textit{Member State} is the main actor of the measures, hence we consider it as an \textit{Agent}. Once the \textit{Agent} has been identified, we have to analyse the sentences for identifying the relations since they make explicit the type of actions that the \textit{Agent} has to undertake.  
In the previous example, we already defined the subject, hence we choose action \textit{Adopt} as the main relation. 

We focus on the following excerpt: \textit{``...the resources required to achieve those objectives, and appropriate policy and regulatory measures, with a view to achieving and maintaining a high level of security...''}. As with many other excerpts, it adds nothing significant to our knowledge because it expresses something that can be considered tautological. In fact, the target of `maintaining a high level of security' is intrinsic in adopting the Directive itself, hence we can decide to skip it. We also avoid to consider words such as ``Shall'' since they only add redundancy. 

By applying the interpretation process to subject   (agent) and verb   (relation), we can deconstruct the sentences of each Article, thereby simplifying their interpretation.

\subsection{Interpreting the NIS 2 objects and complex sentences}\label{sec:method:ob}

At this point, we interpreted that \textit{Member State Adopt} but the object of the sentence is not yet identified. In this section, we show how to identify the object of the sentences and how to deal with complex sentences. 
Usually, in the directive, we can deduce that the scope of one single verb is circumscribed to a sentence. Referring to the excerpt of Section \ref{sec:method:sv}, it is easy to identify as the object of the predicate the phrase ``national cybersecurity strategy''. The final interpreted sentence results in \textit{Member State Adopt National Cybersecurity Strategy}.

In case a security measure does not have to follow the standard structure consisting of subject, predicate and object,  two approaches can be adopted:  a) splitting the sentence into short and simple clauses, b)  considering the part of the sentence following the object to qualify the object itself. The first approach will produce two logical predicates: \textit{Member State Adopt National Cybersecurity Strategy} and \textit{National Cybersecurity Strategy include ...}, whereas the second approach will produce \textit{Member State Adopt National Cybersecurity Strategy} and \textit{National Cybersecurity Strategy} has as qualifying entities \textit{Strategic Objectives} and \textit{Regulatory Measures}, etc. Since those entities are too complex to be considered qualifying entities,  we consider the first approach more adequate. We do not consider the entity \textit{National Cybersecurity Strategy} as an \textit{Agent} since it belongs to a different domain.

\section{Designing an ontology for the NIS 2 directive}\label{sec:onto}
Given the considerations made so far, in this Section, we address the problem of representing NIS 2 by proposing an initial ontology covering articles 7 and 10, termed \niso{}. We adopted Protégé by Stanford University \cite{DBLP:journals/aimatters/Musen15} as IDE for developing the ontology and to acquire the supporting images shown in this paper.
Article 7 was chosen because it is relevant in terms of the definition of the \textit{Member State} agent, while Article 10 provides relations among the agents \textit{Member State} and \textit{CSIRT}.

\subsection{Representation of entities and articles}\label{sec:onto:ea}
\begin{figure}[ht]
    \centering
    \begin{subfigure}{.45\textwidth}
    \centering
       \includegraphics[width=1\linewidth]{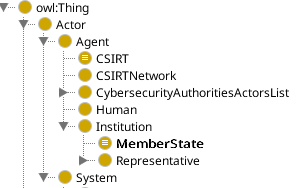}
        \caption{\niso{} hierarchy for classes}
        \label{entities:cl}
\end{subfigure}%
    \begin{subfigure}{.6\textwidth}
    \centering
        \includegraphics[width=.7\linewidth]{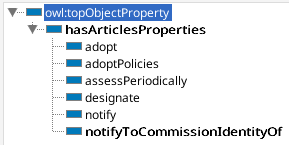}
        \caption{\niso{} hierarchy for properties}
        \label{entities:pr}
    \end{subfigure}

    \caption{A fragment of the NIS 2 Ontology hierarchy.}
    \label{fig:hierarchy}
\end{figure}

We now focus on the semantic representation of articles and related entities through a new ontology. A fragment of the hierarchies for entities and connections is depicted in Figure \ref{fig:hierarchy}. 
Some of the entities such as \textit{Agent} are imported from the ontology \textit{Ontology for Agents, Systems, and Integration of Services},   (OASIS) \cite{oasis, ia2022, woa2022}, to favour its integration with \niso. OASIS is meant to deliver a general representation system and a communication protocol for agents and their interactions, therefore it is profitably leveraged here to inherit the advantages of its \textit{behaviouristic} approach. The behaviouristic approach is an abstraction of operational semantics based on the definition of an agent's behaviour through its decomposition in the essential mental states of the agent, namely, goals and tasks that are sought to be accomplished. Alongside the representation of the agent's mental states, the behaviouristic approach takes care of how behaviours are invoked and put into practice by agents.
The idea of leveraging OASIS is to include a representation mechanism where the behaviours of actors are conveyed. It is useful for a meticulous and controlled placement of the specific actions of the NIS 2 actors because they follow a specific flow: the predetermined verbs formally become \textit{tasks} while the tasks concern specific \textit{objects}. This perfectly adheres to the definitions given above. For example, as an \textit{Agent}, a \textit{Member State} has the task to \textit{notify} the object \textit{National Cybersecurity Strategy}, and so on.

\begin{figure}[ht]
    \centering
    \includegraphics[scale=0.55]{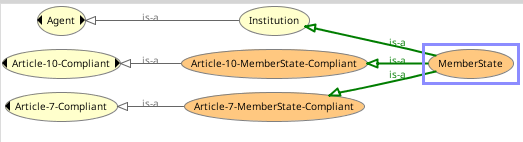}
    \caption{\textit{MemberState} entity specialisation}
    \label{fig:my_label}
\end{figure}

\begin{figure}[ht]
    \centering
    \includegraphics[scale=0.5]{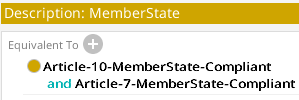}
    \caption{Example of representing the Member State}
    \label{fig:example1}
\end{figure}

In order to group all the articles, we create the class named \textit{NisArticles} and develop the hierarchy according to the level of specialisation. The articles are divided into sub-entities corresponding to the agents by which they are addressed. For example, Article 10 will belong to the hierarchy \textit{NisArticles \textrightarrow  \space Article10 \textrightarrow \space $\{$Article10-MemberState, Article10-CSIRT$\}$}, because the agents involved in such articles are the \textit{MemberState} and \textit{CSIRT} ones, as the hierarchy in Fig. \ref{fig:my_label} depicts. In this way, those subclasses can contain the measures of the specific Article referred to the specific agent.
The tasks related to an agent are not strictly associated with it. As Figure \ref{fig:example1} shows, the tasks that \textit{MemberState} should perform are inherited. We associate the measures of one Article to the corresponding entity, for example, the measures of Article 7 are associated with the \textit{Article7} entity. Then, we associate the \textit{MemberState} entity with the \textit{Article7} entity: this allows us to decouple the security measures from entities that must comply with them. Moreover, the association facilitates  readability in general since the security measures are associated with an entity representing an article and with the actor who is called to comply with those measures.
For example, Figure \ref{fig:example1} shows actor \textit{Member State} to be compliant with the measures of the entities that represent, respectively compliance with Article 7 and Article 10.

\begin{figure}[ht]
    \centering
    \includegraphics[scale=0.5]{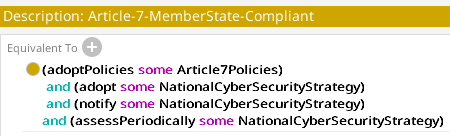}
    \caption{Definition of Article 7 in \niso{}}
    \label{fig:my_label1}
\end{figure}

Following the semantic interpretation described in Section \ref{sec:method}, we represent Article 7 and the related policies as illustrated in Figure \ref{fig:my_label1} in order to guarantee that Article 7 compliant member states verify the related rules, and this can be later verified automatically by reasoning.
Restrictions on the policies could be further qualified if needed to reduce the range of applicability.

Further, we associate qualifying elements to the related entities  through data-type restrictions such as \textit{CyberSecurityIncidentResponsePlan} has \textit{SubmissionMonths} \textit{max 3 xsd:int}.

The integration with OASIS should be considered \textit{working in progress}, as previously mentioned. If on one hand, we have a defined approach for the representation of the NIS 2 directive, fully integrating the corresponding ontology with OASIS requires that it should be perfectly superimposed, for example by associating specific behaviours to agents. That exceeds the scope of this paper and will be explored as future work.

\subsection{Verification of compliance and check of measures}\label{sec:onto:compliance}
The task of compliance verification is complex. It is usually done via auditing between operators, hence causing inconsistencies at times. Using an automatic and time-saver instrument would reduce time and related costs. Ontologies, as mentioned, meet those prerogatives.

Once the previous steps are applied to all security measures, we get a graph with some clusters that identify the most important entities in the directive. In order to validate the verification of compliance mechanism and its results, the last step of our approach is to take into account a generic agent, for example, the \textit{MemberState} entity. 

\begin{figure}[ht]
    \centering
    \begin{subfigure}[t]{.49\textwidth}
    \centering
        \includegraphics[width=1\linewidth]{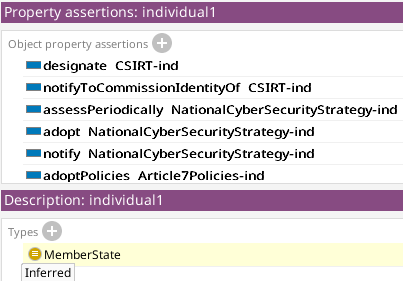}
        \caption{The first test individual}
        \label{entities:ind1}
    \end{subfigure}
    \begin{subfigure}[t]{.5\textwidth}%
    \centering
        \includegraphics[width=1\linewidth]{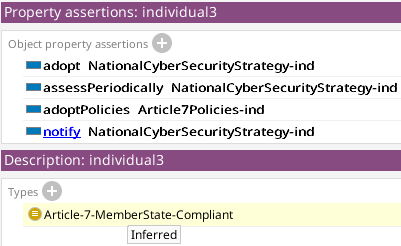}
        \caption{The third test individual}
        \label{entities:ind3}
    \end{subfigure}
    \caption{Test individuals}
    \label{fig:example2}
\end{figure}

We can check its compliance with articles 7 and 10, which currently compose \niso{}.
We created three test individuals   (see Figure \ref{fig:example2}), namely,  \textit{individual1},  which owns all security measures of both articles, \textit{individual2}, which owns no security measure, and \textit{individual3}, which owns only some security measures.    

By running the reasoner, we can see in Figure \ref{fig:example2} the inferences that are carried out. Concerning the first test individual, it is correctly inferred as an instance of \textit{MemberState} since it fulfils all the security measures. Concerning the third test individual, we purposely choose to exclude only the two rules of Article 10, which are \textit{designate CSIRT} and \textit{ensureReportingVulnerabilityTo CSIRT}, hence it is recognised as compliant only with Article 7. Finally, the second test individual has no meaningful inference, because it owns no sufficient rules to derive compliance with both articles.

\section{Conclusions and future works}\label{Conclusions}
We proposed an ontological approach for the characterisation of security directives. The approach allowed us to propose a structural solution for translating security documents to a mathematically-driven world. The paper proposes just an approach then proper methodologies and principles for the ontological design will be later addressed deeply.
The target of the work was the NIS 2 directive but analogous considerations can be made for similar directives. 
As a result, we defined an ontology termed \niso{}, which currently covers articles 7 and 10 of the NIS 2. While this is clearly incomplete with respect to the whole directive, it provides a paradigmatic representation of some of the essential entities and relations that the directive stands upon. 
Our approach meets the \textit{FAIR} principles and the \niso{} may help security analysts to quickly verify the status of the institutions' compliance, resulting in an efficient search engine for security measures. 

The target of future works is to expand the prototype to the entire directive and to refine the correlation with the OASIS ontology.
Our ontology, once completed, could be integrated with ``Towards Unified European Cyber Incident and Crisis Management Ontology'' proposed by ENISA \cite{articleEnisa}. That would be possible since both revolve around common entities, such as \textit{Member State} and \textit{threat}.

\bibliographystyle{unsrtnat}
\bibliography{references}
\end{document}